\documentclass[twocolumn]{article}
\usepackage[numbers]{natbib}
\usepackage{color,soul}
\usepackage{graphicx}
\usepackage[bf]{subfigure}% http://ctan.org/pkg/subfigure
\usepackage{siunitx}
\usepackage{ragged2e}
\usepackage{float}
\usepackage{placeins}
\usepackage{framed}
\usepackage{multicol}
\usepackage{xcolor}
\usepackage{varwidth}
\usepackage{booktabs}
\usepackage{bbding}
\usepackage{pifont}
\usepackage{wasysym}
\usepackage{amssymb}
\usepackage{siunitx}
\usepackage[switch]{lineno} 
\usepackage{lipsum} %Creates example text

\usepackage{color, colortbl}
\usepackage[font=bf]{caption}
\DeclareCaptionLabelSeparator{divide}{ $|$ }
\usepackage{bibunits}
\usepackage{hyperref}
\hypersetup{
    colorlinks=true,
    linkcolor=blue,
    citecolor=black,
    filecolor=magenta,      
    urlcolor=cyan,
    pdftitle={Overleaf Example},
    pdfpagemode=FullScreen,
    }
    
\usepackage{etoolbox}
\usepackage{soul}
\defaultbibliography{cas-refs}
\defaultbibliographystyle{vancouver}
\makeatletter
\newcommand*{\newbibstartnumber}[1]{%
  \apptocmd{\thebibliography}{%
    \global\c@NAT@ctr #1\relax
    \addtocounter{NAT@ctr}{-1}%
  }{}{}%
}
\makeatother

\usepackage{lipsum}

\begin{document}

%\linenumbers
%\lipsum[1-20]

\author{
Jing Han$^{1*}$ \and 
Tong Xia$^{1*}$ \and 
Dimitris Spathis$^{1}$ \and 
Erika Bondareva$^{1\dagger}$ \and 
Chloë Brown$^{1\dagger}$ \and
Jagmohan Chauhan$^{1,4\dagger}$ \and
Ting Dang$^{1\dagger}$ \and
Andreas Grammenos$^{1\dagger}$ \and
Apinan Hasthanasombat$^{1\dagger}$ \and
Andres Floto$^2$ \and
Pietro Cicuta$^3$ \and
Cecilia Mascolo$^1$ \and
}

\date{%
    $^1$Department of Computer Science and Technology, University of Cambridge, UK\\%
    $^2$Department of Medicine, University of Cambridge, UK\\%
    $^3$Department of Physics, University of Cambridge, UK\\%
    $^4$ECS, University of Southampton, UK\\%

    \small{$^{*}$JH, TX contributed equally to this study}\\%
    \small{$^{\dagger}$listed in alphabetical order, contributed equally}
    \\[2ex]%
    %\today
}

\title{\textbf{Sounds of COVID-19: exploring realistic \\ performance of audio-based digital testing}
% Sounds of COVID--19:\\ from digital screening to disease progression}
% \tx{Listening to COVID-19: a digital test, a reliable evaluation, and a holistic study}
% \jh{COVID-19 Sounds: development and validation of an audio based COVID-19 prediction model}
% %\cm{Can we trust the sounds of COVID-19 for digital testing?} \ds{I would avoid titles with questions ;) (see Betteridge's law) }
% \cm{Sounds of COVID-19: exploring realistic performance of audio based digital testing}
} 

\maketitle

Researchers have been battling with the question of how we can identify Coronavirus disease (COVID-19) cases efficiently, affordably and at scale.
Recent work has shown how audio based approaches, which collect respiratory audio data (cough, breathing and voice) can be used for testing, however there is a lack of exploration of how biases and methodological decisions impact these tools' performance in practice.
In this paper, we explore the realistic performance of audio-based digital testing of COVID-19. 
To investigate this, we collected a large crowdsourced respiratory audio dataset through a mobile app, alongside recent COVID-19 test result and symptoms intended as a ground truth.
Within the collected dataset, we selected 5,240 samples from 2,478 participants and split them into different participant-independent sets for model development and validation. Among these, we controlled for potential confounding factors (such as demographics and language).
The unbiased model takes features extracted from breathing, coughs, and voice signals as predictors and yields an AUC-ROC of 0.71 (95\% CI: 0.65$-$0.77).
We further explore different unbalanced distributions to show how biases and participant splits affect performance. 
Finally, we discuss how the realistic model presented could be integrated in clinical practice to realize continuous, ubiquitous, sustainable and affordable testing at population scale.

\section*{Introduction}

% \jh{revised}
\label{introduction}
% \begin{itemize}
%     \item scientific and clinical background, including the intended use and clinical role of the index test
%     \item study objectives and hypotheses
% \end{itemize}    
%\cm{find right template}
% https://coronavirus.jhu.edu/map.html

Since its outbreak in early December 2019, over 169 million cases of the novel coronavirus disease have been reported, including 3.5 million deaths. 
Researchers and scientists have made considerable strides in developing treatments and vaccines for COVID-19, and effective and easily accessible tests have been key to trace infected people quickly. Currently, the most commonly used and first-line diagnostic tool for COVID-19 is the reverse transcription polymerase chain reaction (RT-PCR) assay to detect the presence of viral ribonucleic acid (RNA) from swab samples~\cite{Cevik20-Virology, Vogels20-Analytical}. 
RT-PCR tests are highly sensitive in laboratory testing (over 95\% diagnostic sensitivity and specificity), however they have been found to perform differently in several commercial kits, with sensitivity ranging from 75\% to 100\%, and in the worst case reaching as low as 38\%~\cite{Garg20-Evaluation, Liu20-Positive, Van20-Comparison}.
Moreover, the sample analysis process is involved, time-consuming, and limited to approved laboratories with highly-trained staff, leading to limited testing capacity and failing to meet the rapid increase in demand. 
It is crucial that the pandemic response overcomes these challenges to timely test on a massive scale. This requires fast, affordable, sustainable and effective testing methods, which can be repeated over time by individuals to track progression. This would help contain the current spread but also suppress resurgence and minimise health risks.

Within this context, in the past year researchers have developed and published multiple models for COVID-19 prediction using audio~\cite{Menni20-Real, mei2020artificial,Harmon20-Artificial, javaheri2020covidctnet, xu2021ai, lee2021deep}. In particular, advances in machine learning have demonstrated the potential of automated auscultation of respiratory sounds and brought about new possibilities for fully automated COVID-19 screening~\cite{Imran20-ai4covid, Brown20-Exploring, laguarta2020covid, pinkas2020sars, Han2021exploring, 9361107, coppock2021end, 9414201}. For instance, a systematic review by Wynants et al reports that AUC-ROC (Area Under the Receiver Operating Characteristics) performance
%\cm{do we need to spell out auc roc somewhere?} 
of over 75 existing COVID-19 prediction models~\cite{wynants2020prediction} is in the range of .70 and .99.

There is however a lack of studies exploring the biases and model evaluation processes which affect (potentially even positively but unrealistically) these performance results.
%Unfortunately, all previous studies have 
% \tx{the real deployment} 
%limitations to be deployed in real life, which are mainly due to the nature of the data collected and the machine learning processes applied to it.
% \st{Examples include biases\cm{cite and explain}, non representative data \cm{cite and explain}, model overfitting\cm{cite and explain}. }
% \cm{agree with sally: this paragraph is very important and we need to be precise} 
% For example, \tx{I could suggest we use bullets 1,2,3 to make the discussion clearer} 
Such issues include:
\begin{itemize}
    \item Potential underlying data biases or study limitations  not reported sufficiently, where models were developed and evaluated with limited data which might not be representative of the target population (e.g., 19 subjects in~\cite{9414201}, 51 subjects in~\cite{Han2020}, and 88 subjects in~\cite{pinkas2020sars}). 
    \item Risk of model overfitting, especially when deploying complex modelling strategies (e.g., a 100\% accurate diagnosis of asymptomatic COVID-19 individuals was reported in~\cite{laguarta2020covid}).
    % \tx{Can other understand what we want to say by 'overfitting'? shall we say lack of transparent and re-productivity? }. 
    \item Methodological flaws (e.g. same users during model development and validation~\cite{xue2021exploring}) which would be unrealistic in a practical clinical setting, resulting in artificial performance boost. 
    \item Lack of systematic comparison with other  respiratory diseases like asthma and bronchitis, and only distinguish COVID-19 from healthy controls~\cite{subirana2020hi}. 
\end{itemize}
% Similarly, 
% Moreover, 
% Additionally, 
Due to these issues, many researchers raised concerns on the feasibility and effectiveness of such models if deployed in real settings~\cite{wynants2020prediction, topol2020my, roberts2021common}.
%\jc{when or if. are they already deployed or it is still a possibility}when being deployed in real life

%\textbf{Does this mean audio based COVID-19 testing looks promising on the papers but might not in work effectively in the real world?}
% is unreliable and should not be used \tx{promising in the papers but not in real use :)}? 
%have issues and flaws in design and/or reporting, such as the high risk of bias\td{in terms of what?}, inappropriate and unrepresentative data\td{regarding ?}, model overfitting, and vague reporting.\jh{which paper(s) to site, or not to cite at all?} \tx{Not sure overfitting is suitable here, we have this issue due to the limited data}\td{This reads a bit vague to me, and might not make our motivations in the next paragraph clear.Maybe the statement of the flaws can be a bit specific and clear, which can also correspond to our work in the next paragraph.} 
% The systematic review demonstrated that, the performance of these identified models is probably optimistic and none can currently be recommended for clinical use~\cite{wynants2020prediction}.\cm{I do not understand the meaning of this sentence but i left it here}

%Thus, there is an urgent need to develop and validate the effectiveness and generalisation of the COVID-19 prediction model to simulate being deployed in a realistic scenario.

In this work, we investigate the limits of audio-based COVID-19 testing with the aim of creating the foundation of {\em realistically applicable} audio tools.
The aim of this study is two-folds: First, based on a large crowdsourced dataset, to investigate the performance of an audio-based testing method when working with, to the best of our knowledge, unbiased data with a methodological design based on realistic assumptions (e.g. independent user split). Secondly, to explore the impact of biases and design pipeline on performance.
%The aim of this study is data, carefully design and develop a sound-based COVID-19 prediction model, responsibly validate its capability for various target populations, and comprehensively discuss its potential implications for clinical practise in the light of the findings of the study. 
% In this paper, we develop a deep learning (DL) model for COVID-19 screening, by automatically and jointly analysing three types of respiratory sounds (i.e., breathing, cough, and voice). 

For this purpose, we first gathered crowd-sourced respiratory sound data from general population via smartphones. We carefully prepared data for model development and validation, by selecting representative audio samples from self-declared COVID-19 positive or negative participants.
% \cm{unclear what positive and negative cases are, not epxlained}
Subsequently, we developed a deep learning model on a portion of the data and then validated its predictive performance on an independent population. In particular, we adhered to the TRIPOD reporting guideline~\cite{moons2015transparent}, aiming at reporting in a complete, transparent and usable manner. Our discussion explores the biases and how machine learning model hyperparameters could be tuned, depending on the use of the tool (e.g. on symptomatic or asymptomatic populations) and public health needs.

\section*{Results}\label{results}
% \begin{itemize}
    % \item Data preparation and statistics \tx{re-wrote}
    % \item Model overall performance on balanced test set\tx{re-wrote}
    % \item Model performance on varied prevalence rate\tx{re-wrote}
    % \item Model performance on various health status and smoking  status \tx{re-wrote}
    % \item performance comparison with human doctors \jh{to add}
% \end{itemize}

\begin{figure*}%[pos=t,width=3cm,align=\centering]
\centering
{\includegraphics[width=1\textwidth]{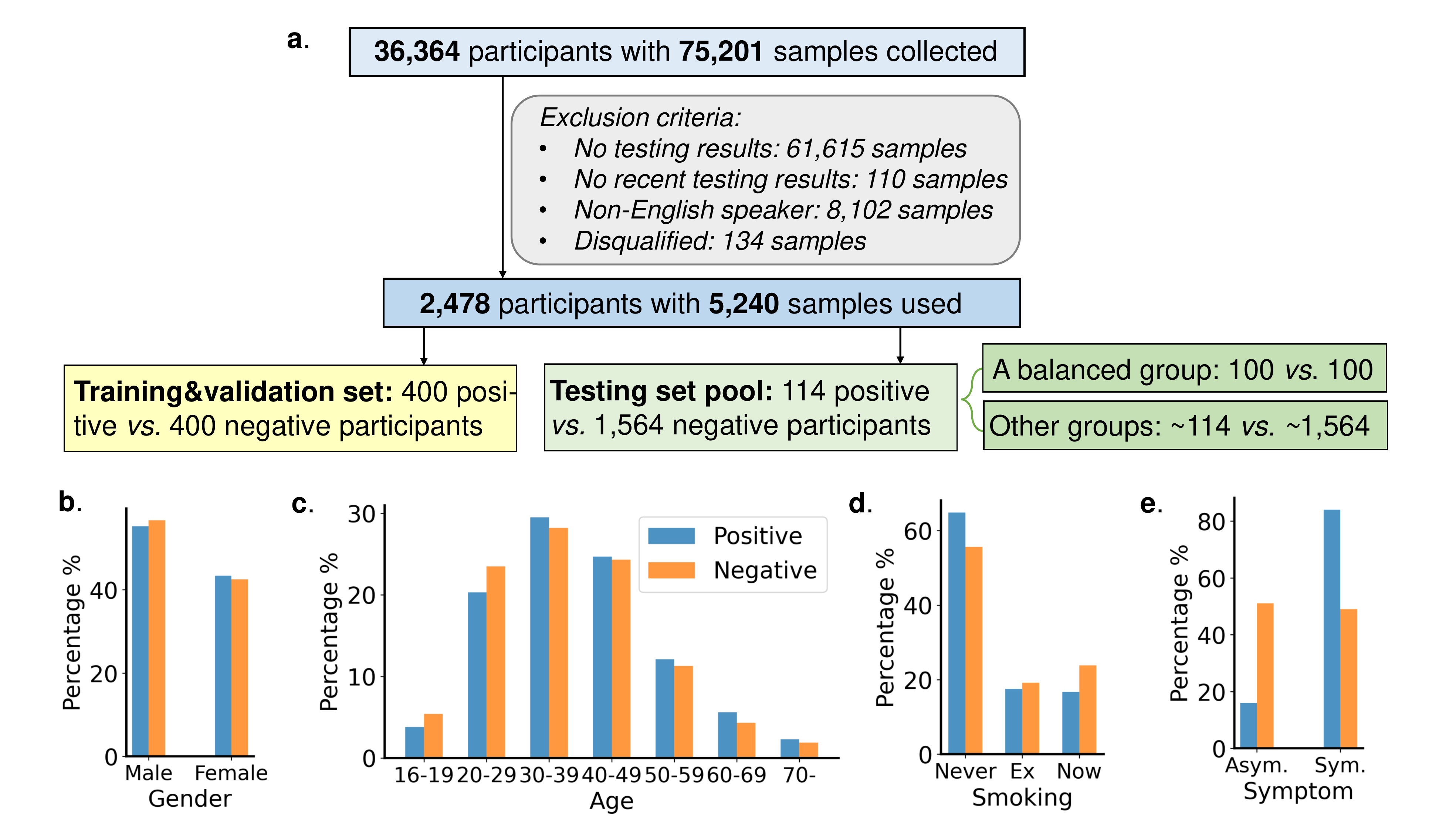}}
\caption{\textbf{Data flow diagram and demographic statistics.} \textnormal{\textbf{a)} Data cleaning and selecting with 514 COVID-19 positive and 1,964 COVID-19 negative participants eventually included for experiments.  \textbf{b-e)} Demographic statistics for the included 2478 participants with blue representing positive class and orange -- negative class:  \textbf{b)} Gender distribution with about 56\% male, 43\% female and 1\% preferring not to say both two classes. \textbf{c)} Age distribution with more than 87\% positive and negative participants aged 20-59. \textbf{d)} Smoking history with more than half participants never smoking and the remaining participants having smoked before or smoking currently. \textbf{e)} symptom distribution: 16\% of the positive group showed no symptoms, while 49\% of the negative group reported at least one of the symptoms. } } 
% \jh{TODO:update new figures, without preliminary task, nor longitudinal analysis} \tx{Done} \ds{Replace Germany with DE instead of GE.}
\label{fig:data}
\end{figure*}
\begin{figure*}[t]%[pos=t,width=3cm,align=\centering]
\centering
{\includegraphics[width=0.99\textwidth]{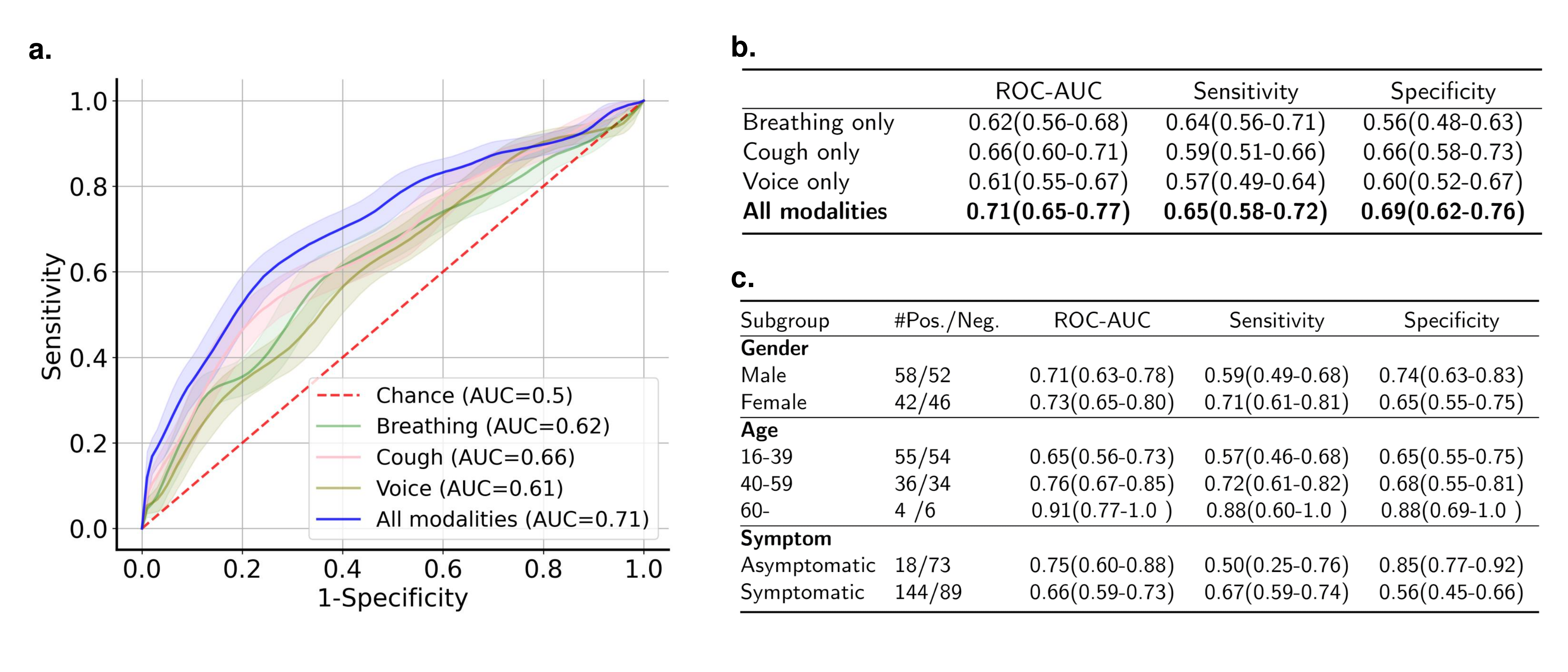}}
\caption{\textbf{Model performance}. \textnormal{\textbf{a)} Receiver-operating characteristic curve for the binary classification task of diagnosing COVID-19. \textbf{b)} ROC-AUC, sensitivity and specificity with 95\% confidence intervals in brackets 
% \td{spell out?} \tx{specify what?} 
for combination of all modalities or each single modality separately. \textbf{c)} Subgroup performance comparison under three modalities. For gender and age group, $\#$ denotes the number of unique positive/negative participants. Note that some participants provided multiple samples, which could be either asymptomatic or symptomatic.}}
% \jh{to be updated}\tx{Done}\ds{The yellow line doesn't read so well here. Maybe change?}
\label{fig:results}
\end{figure*}

\noindent\textbf{Dataset preparation and statistics.} 
For data gathering purposes, an app\footnote{https://www.covid-19-sounds.org/en/} was developed and released in April 2020 to crowd-source participants' demographics, medical history, symptoms, COVID-19 test results and  audio recordings: three voluntary cough sounds, three to five breathing sounds, and three speech recordings where user was asked to read a specific sentence. The self-reported COVID-19 test results included receiving a positive/negative test result, or not being tested before the recording. More details can be found in Methods in \hyperlink{Appendix}{Appendix}. By the end of April 2021, more than 36,000 participants contributed approximately 75,000 samples. Around 5,600 participants submitted at least one sample with COVID-19 positive or negative result within 14 days before the recording, with a total of 2,080 positive samples. Most participants came from United States, United Kingdom and other European countries. 

%\jc{the following sentence is unclear}As COVID-19 prevalence levels from \jc{what is those} those countries in our dataset were different, language might be a confounding but not useful factor for COVID-19 prediction, particularly in terms of voice.   For instance, the percentage of positive participants who speak \jc{please change to Italian} Italy was much higher than that of English. Without data selection and control, the learned model is likely to capture language as a character to predict COVID-19.
Our app is a multi-language tool, but in this study we focus only on audio samples from English-speaking participants (77.7\% of the overall participants) to avoid language-related bias. Audio quality checks were conducted to filter out incomplete or noisy samples. Finally,  2,478 participants (514 positive and 1,964 negative participants) with 5,240 samples were included for experiments, as shown in Fig.~\ref{fig:data}a (more detailed data selecting criteria can be found in  \hyperlink{methodSection}{Methods}). Demographic statistics for the experimental data are presented in Fig.~\ref{fig:data}b-d: 56\% participants in the selected data were male, the majority aged 20-49, half never smoked. In addition, as shown in Fig.~\ref{fig:data}e, 84\% of the participants who tested positive reported symptoms like fever or cough, while others did not report any symptoms at the recording time. 51\% of the negative participants reported no symptoms, while 49\% had symptoms such as dry/wet cough, fever, dizziness, etc. 
%These distributions demonstrate the representatives of our collected data for a realistic evaluate.

%\cm{shall we talk more about this bias earlier perhaps given we clearly do something about it?} \tx{Rephrased, please check}
%\cm{do we explain how we get this testing positive or negative, it's important we say it's crowdsourced}\tx{added}
%\cm{do we say we ask them to report med history and symptoms?} \tx{added}

From the 2,478 English-speaking participants, we prepared a training $\&$ validation  set which consists of 800 participants with balanced COVID-19 status and other demographics to optimise the parameters of our deep learning model, as labelled by the yellow box in Fig.~\ref{fig:data}a, and the rest of the data were used for evaluation, namely testing set pool (green box in Fig.~\ref{fig:data}a).
To inspect the performance in different realistic deployment scenarios, we first held out a representative testing set with balanced and demographic controlled participants, containing 100 positive and 100 negative participants.
 Furthermore, we randomly selected positive and negative participants from the testing pool to form new groups with the criteria of various prevalence levels, medical history, smoking status, to holistically validate our model. 
Apart from the controlled training and testing data, to simulate the impact the unrealistic experiment setting and bias, we also prepared some training and testing sets with improper data splitting and various biases the model. Details can be found in \hyperlink{methodSection}{Methods}. Accuracy is presented by ROC-AUC (Area Under the Receiver Operating Characteristics curve), sensitivity and specificity. 

\begin{figure*}[t]
  \centering
  \includegraphics[width=0.99\textwidth]{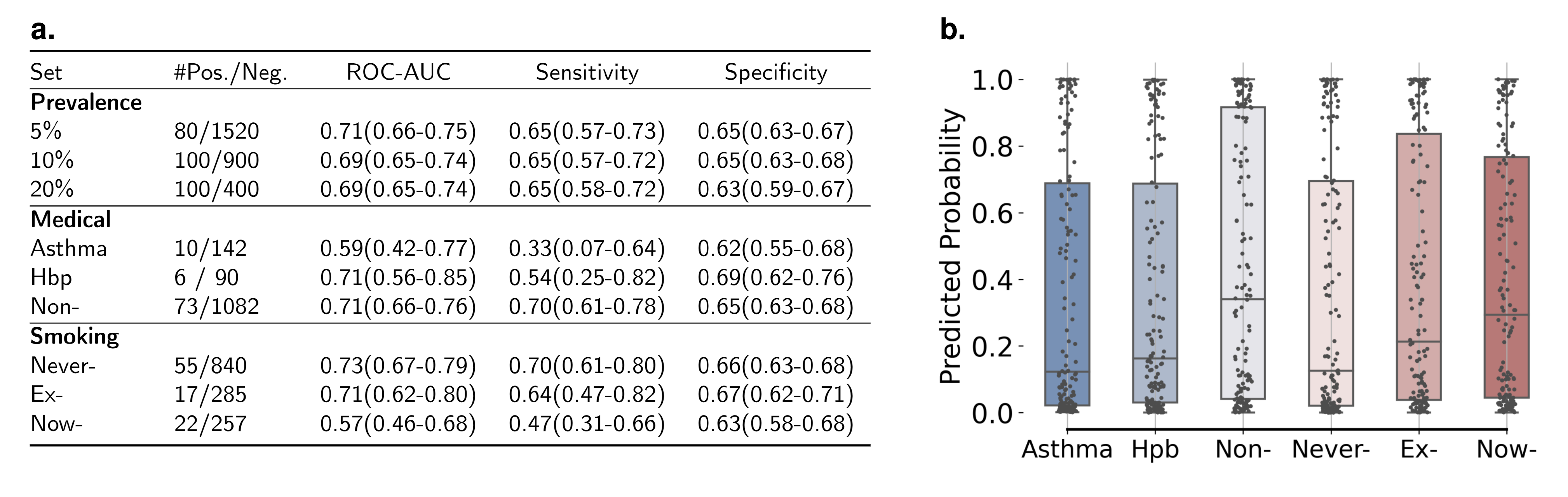}
  \caption{Performance comparison under different conditions. \textnormal{\textbf{a)} Performance for different prevalence levels, for participants with asthma/HBP or without any medical history, and for never-, ex- and now-smoking participants. $\#$ denotes the number of positive/negative participants. \textbf{b)} Predicted probabilities of infecting COVID-19 on those negative participants,  with values above 0.5 indicating a false positive. No significant difference was observed across subgroups.}}
  %\eb{this is a little convoluted, perhaps "Probability for COVID-19 negative participants that the sample is a false negative"? if i understood this correctly}\tx{It is the predicted probability of getting COVID for all negative users, if the prob$>$0.5, it is false positive.}\eb{thanks for the explanation, this makes sense. And I guess there isn't really a clearer way of wording this. Perhaps could say instead of "Positive probability" -- "Probability that the participant has COVID-19 for users who tested negative, with values above 0.5 indicating a false positive -- no significant difference was observed across subgroups." not sure if my suggestion is better, so feel free to either use it or keep whatever we have now}}}
  \label{fig:Tas2Others}  
\end{figure*}

%\cm{do we say why balanced is important?} \tx{It is one of the testing case, as important as different prevalence. in addition to the same number of positive and negative, demographics are also controlled for subgroup study}
%\jc{Please find a better heading}
\vspace{.2cm}\noindent\textbf{COVID-19 detection performance.}
On the demographic-representative testing set with 200 participants (see age and gender distribution in \hyperlink{Appendix}{Appendix} Table~\ref{tab:data}), 
%On the balanced testing set with 200 participants, \eb{I see this was something raised before, but I also at first was struggling to understand where 200 came from. Perhaps worth adding in the previous subsection a few more details on the splitting of the testing set into balanced and unbalanced subsets, because currently its a little difficult to follow}
%\cm{unclear why from 800 to 200} \tx{800 is the training sieze}
 our deep learning model with three sound types yielded a ROC-AUC of 0.71 (0.65-0.77) (Fig.~\ref{fig:results}a), with sensitivity 0.65 (0.58-0.72) and specificity 0.69 (0.62-0.76)  (Fig.~\ref{fig:results}b). The combination of three sound types outperformed any single modality: a ROC-AUC of 0.66 (0.60-0.71) on cough, 0.62 (0.56-0.68) on breathing, and 0.61 (0.55-0.67) on voice was achieved. Moreover, breathing yielded the highest sensitivity of 0.64 (0.56-0.71), but cough showed the highest specificity of 0.66 (0.58-0.73). This indicates that all modalities are informative, and their combination leads to the optimal performance.
We further tested the performance of the model on different demographic subgroups under this testing set (Fig.~\ref{fig:results}c), which showes similar results across different age and gender distributions: ROC-AUCs were all  above 0.65, and sensitivity and specificity were similar for each group. Accuracy on the over-60 subgroup is slightly higher, but we suspect that the increased performance might be a result of the limited number of participants in this group.
%The insignificant variation indicates the robustness of our model and unbiased COVID-19 prediction capability.
%\td{It seems that the metrics is significantly higher for users ageing above 60 of 0.91 compared to 0.7ish. This might need one sentence to clarify the reason? Is there any evidence showing that aged users might get severe affection that the voice bio-markers are more distinguishable?} \tx{since we only have 4+6 over60 users, it is not convincing the AUC of 0.91 is significant. We might not want to emphasise it ?}\cm{fair enough}
%\jc{sorry it is not insignificant} 
We also inspected how symptoms impact the model performance by dividing the 200 participants in the testing set into asymptomatic and symptomatic subgroups. As presented in Fig.~\ref{fig:results}c, for both subgroups our model yielded ROC-AUCs above 0.66. Yet, from the comparison we can also observe that our model performs better on distinguishing asymptomatic negative participants (specificity=0.85 (0.77-0.92)) and symptomatic positive participants (sensitivity=0.67 (0.59-0.74)). While for more challenging cases, i.e. predicting symptomatic negative and asymptomatic positive cases, we achieved a lower accuracy: Sensitivity of 0.50 (0.25-0.76) for asymptomatic and specificity of 0.56 (0.45-0.66) for symptomatic participants. A potential explanation could be that asymptomatic positive participants might not manifest changes in audio characteristics, and thus are intractable for detection. Further discussion about the real implications and applications can be found in \hyperlink{discussionSection}{Discussion}.

\noindent\textbf{Model performance on varied prevalence rate.}
According to a recent statistical study, the prevalence rate of COVID-19 ranges from 0.12\% to 33\% worldwide~\cite{louca2020covid}. Therefore, in addition to testing our model in a balanced setting (50\% prevalence level, Fig.~\ref{fig:results}), we evaluated its performance in various prevalence scenarios. To simulate this, we re-sampled participants from the testing pool (Fig.~\ref{fig:data}a) and lowered the proportion of COVID-19 positives to 5\%, 10\% and 20\% (Fig.~\ref{fig:Tas2Others}a). The performance does not degrade compared to that of 50\% prevalence (Fig.~\ref{fig:results}b): ROC-AUC of 0.71 (95\% CI 0.66-0.75), 0.69 (0.65-0.74), and 0.69 (0.65-0.74) can be achieved on 5\%, 10\%, and 20\% prevalence levels, respectively.  
%\jc{any reason we did not try 30\%} 
This is a promising result, suggesting the potential of AI-enabled COVID-19 screening in the real world.

\vspace{.2cm}\noindent\textbf{Model performance on various health and smoking status.}
One of the most important concerns to address is whether these audio models for COVID-19 testing might be confusing COVID-19 with other illnesses or respiratory pathologies. To investigate this topic further, we split the participants of the testing pool into several non-overlapping groups. Results are presented in Fig.~\ref{fig:Tas2Others}.

The first controlled criterion is medical history. We selected participants who reported that they have \textit{Asthma}, \textit{HPB (High Blood Pressure)}, and those who claimed having no medical history. We compared the model performance and found that all metrics reach comparable level of accuracy on average: the specificity on Asthma group was 0.62 (0.55-0.68), on HPB group was  0.69 (0.62-0.76), on no medical history group was 0.65 (0.62-0.76) (Fig.~\ref{fig:Tas2Others}a), and on a mix of participants was 0.69 (0.62-0.76) (Fig.~\ref{fig:results}b).
Predicted COVID-19 probabilities of the negative participants based on our model are compared in Fig.~\ref{fig:Tas2Others}b. Kruskal-Wallis H Test~\cite{mckight2010kruskal} on the three negative groups' probabilities yielded a p-value of 0.62 ($>$0.05), showing that the predictions are from the same distribution. This validated the assumption that the medical history cannot confuse our model. Worth noting, that the declined sensitivity for \textit{Asthma} and \textit{HPB} groups might be caused by the very limited number of testing samples, leading to relatively large performance fluctuations. 
%A plausible reason for the declined sensitivity for \textit{Asthma} and \textit{HPB} groups can be the very limited testing samples, leading to relatively large performance fluctuations. 

\begin{figure*}[!t]
\centering
{\includegraphics[width=0.95\textwidth]{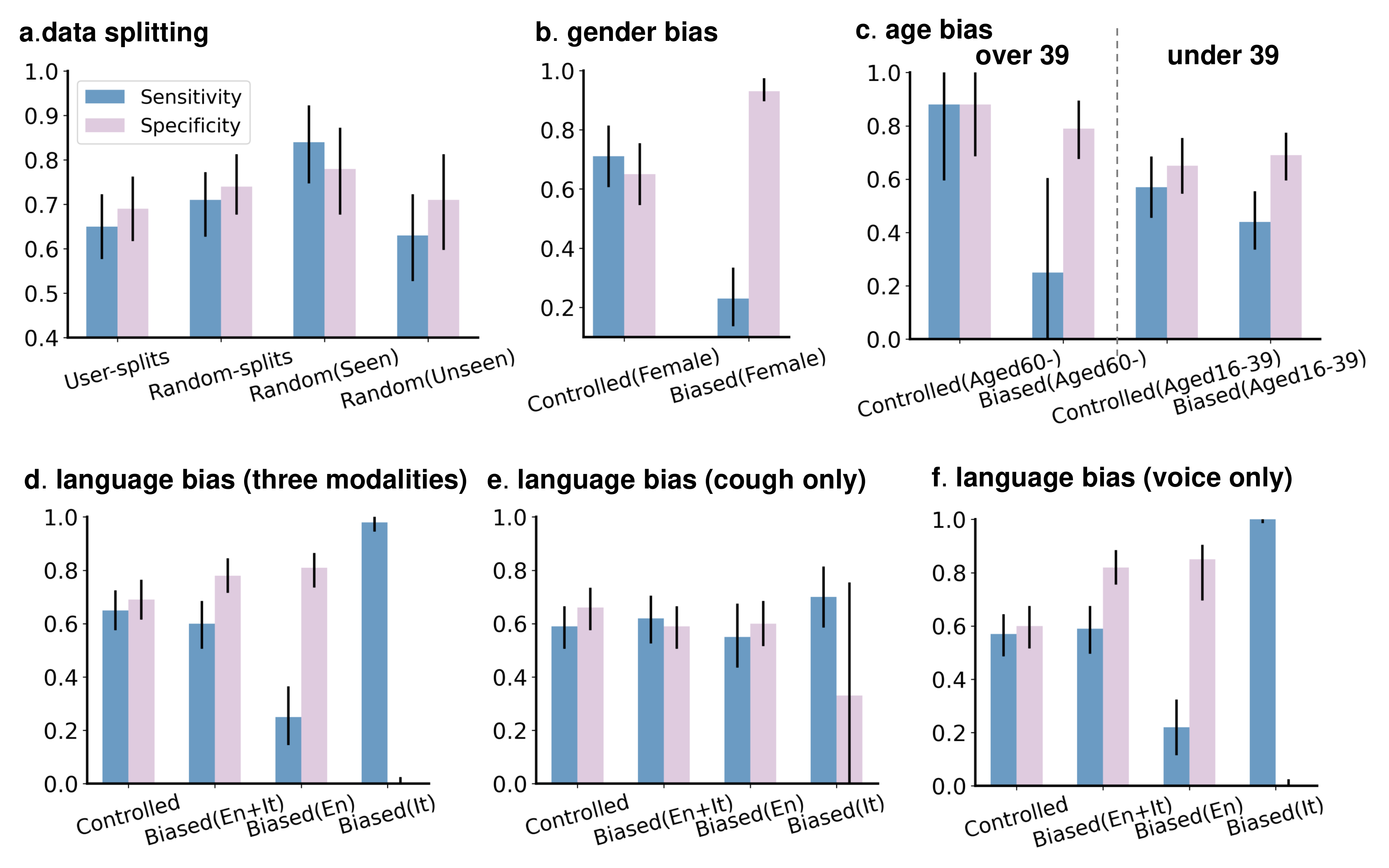}}
\vspace{-0.05cm}
\caption{Performance comparison. \textnormal{Sensitivity (blue) and specificity (pink) are presented with sold liner showing the 95\% CIs. If not particularly mentioned, the results are based on the combination of three sound types.}
a, User-independent splits \textit{v.s.} sample-level random splits: \textnormal{(Seen) denotes the performance on samples whose other samples were used for training, otherwise the performance is notated by (Unseen).} 
b, Controlled demographics  \textit{v.s.} gender bias: \textnormal{(Female) denotes the female subgroup.}
c, Controlled demographics \textit{v.s.} two types of gender biases: \textnormal{all negative participants in training set aged over 39 or under 39. (Aged60-) and (Ages16-39) denote the elder and the younger subgroup.}
d, e, and f, Model for English-speakers  \textit{v.s.} model for biased English- and Italian-speakers: \textnormal{(En) and (It) denote two subgroups from the testing set.}
% \tx{How to put the figure before method?}
}
\label{fig:results_bias}
\end{figure*}

The second controlled criterion is the reported \textit{smoking} status. The variance of the performance across groups was marginal (Fig.~\ref{fig:Tas2Others}a): Specificity for those never smoking was 0.66 (0.63-0.68), for those having quit smoking was 0.67 (0.62-0.71), and for those smoking currently was 0.63 (0.58-0.68). Similar to medical history, predicted probabilities for these three are presented in  Fig.~\ref{fig:Tas2Others}b, with a p-value of 0.51 ($>$0.05) from Kruskal-Wallis H Test. Sensitivity for smokers was slightly lower: 0.47 (0.31-0.66), which might be explained by the fact that five of the 22 COVID-19-positive smokers were asymptomatic (23\% in this group against 16\% in Fig.~\ref{fig:data}e). As our model is  better in predicting symptomatic COVID-19 correctly, this explains the slight drop in the overall sensitivity for this group.  
%\jc{unclear sentence}Overall, our model is able to distinguish the impact of COVID-19 on sounds from the impact of other potential factors. 

% \jh{Shall we point out the low sensitivity results for asthma and now-smoking participants?} \tx{If we have to explain, i think it mainly because less than ten users are included in the positive group, leading to a wide CI and low average SE. }\td{From the numbers it seems a drop in asthma and now-smoking group in AUC and sensitivity, and from the figure the non-med-history seems to be different from other two conditions as well, with larger mean and standard deviation. I am not sure whether the claim of no difference can hold here? Or is there any metrics or statistical significance that we can show here? Or any explainable reason which I guess limited number of users could be the key one?} \tx{AUC might not be a fair metric to compare as the number of positive/negative are different. For fig3b, yes the mean is not the same, is a p value showing the insignificant difference a good evidence here?  }
% \cm{we should say something one way or the other i think..}
% \tx{A h test result was supplemented with p>0.05. So the difference is insignificant. Please check this paragraph again. }

\vspace{.2cm}\noindent\textbf{Model performance with unrealistic evaluation and biases.} To show how the bias and unrealistic experiment design impacts the model performance, we re-selected and purposefully introduced various biases that previous works might have had, to generate another four training and testing sets to attempt to artificially inflate the results.
%\jc{i hope we are reanalysing our results on our simulated dataset here and not reproducing other people's results}\tx{yes, 're-produce' is not good, I rephrased.} 
The artificially created biases are as follows: 1) Using sample-level random splits (random-splits for short) instead of participant-independent splits (user-splits for short) for training and testing. 2) Introducing gender bias into the data by selecting 85\% of the negative participants as female. 3) Bringing age bias into negative group. There are two biased groups: selecting all negative participants as who aged over 39 (Group 1) and as who aged under 39 (Group 2).  4) Replacing some English-speaking participants with Italian-speaking participants and making the proportion of Italian-speaking participants relatively higher in the positive group.
Details of the data used for comparison can be found in \hyperlink{methodSection}{Methods} and \hyperlink{Appendix}{Appendix} Fig.~\ref{fig:results_age}-\ref{fig:results_lan}. We trained the model without changing the network structure, and if not specifically mentioned, the results are based on the combination of three sound types (breathing, cough and voice).

%\eb{need to reread this later, overall a little complex and very long}
Fig.~\ref{fig:results_bias} presents the key findings (a detailed comparison can be found in \hyperlink{Appendix}{Appendix} Table~\ref{tab:UserBias}-\ref{fig:results_lan}). From Fig.~\ref{fig:results_bias}a, random-splits yielded a higher accuracy than user-splits, with the performance gains coming from the overlapping participants whose data have been \textit{seen} from training: with sensitivity of 0.84 (0.75-0.92) and specificity of 0.78 (0.68-0.87), since personal sound traits are easy to memorise for the model. However, this is less realistic, as in real world scenarios the model should ideally be well adapted to unseen new population. This also may validate our hypothesis that some previous works reported optimistic performance  by using this random-split protocol. %\eb{over-optimistic isnt good. Should say optimistic, or inflated, or overfitting.. perhaps "that some previously reported results might have been sufferring from overfitting"}
Demographic bias either in age or gender appears to also lead to biased results. Overall ROC-AUC might be boosted as shown in \hyperlink{Appendix}{Appendix} Table~\ref{tab:GenderBias}-\ref{fig:results_age},  but a great difference between sensitivity and specificity can be observed in some subgroups. For instance, sensitivity of 0.23 (0.14-0.33) but specificity of 0.93 (0.90-0.97) were obtained as shown in Fig.~\ref{fig:results_bias}b on biased (Female) group, because positive females were under-represented in the training set and this model tends to treat female participants as negative. Similar results can be observed in age-biased groups (see Fig.~\ref{fig:results_bias}c). In the group where negative participants aged-over-39 in training set, the model yielded higher specificity than sensitivity on the aged-over-60 participants in the testing set. On the contrary, the model trained from the data biased to aged-under-39 negative participants yielded higher specificity on the younger group (seen Fig.~\ref{fig:results_bias}c).  When it comes to the language bias, i.\,e., for Italian-speakers, positive participants were over-represented for training, we get the results that sensitivity is as low as 0.25 (0.15-0.36) in English subgroup and specificity is close to 0 in Italian subgroup from Fig.~\ref{fig:results_bias}d, and this bias particularly impacted voice modality (see Fig.~\ref{fig:results_bias}f) and slightly influenced cough (see Fig.~\ref{fig:results_bias}e).
Yet, our performance (namely controlled model in Fig.~\ref{fig:results_bias}) shows consistent sensitivity and specificity across all subgroups, presenting a realistic value for model application.

\section*{\hypertarget{discussionSection}{Discussion} }\label{discussion}
% \begin{itemize}
%     % \item monitor progression?
%     \item clinical implication: use scenario? asymptomatic screening/ symptomatic diagnosis? \tx{To be added}
%     \item limitations 
%     \item future efforts/plans
% \end{itemize}

\vspace{.2cm}\noindent\textbf{Comparison with other studies.} 
% \textit{Importance of controlling potential confounding} 
% \tx{emphasise more how our study differs from previous work in this aspect?}\jh{done} 
For digital technologies to penetrate the clinical practice it is pivotal that studies become more explainable and that the models are resilient to the data noise, variability and bias present in real data.

{\em Demographic Bias}: In our study design and data selection we concerned ourselves with potential confounding factors and tried to rule out selection bias, as these may lead to unrealistically inflated results. Specifically, we split positive and negative samples into three partitions for model training, optimisation, and testing, while adjusting the data partitioning and maintaining similar distributions of age and gender across different data splits to control for potential confounding variables (see Appendix Table~\ref{tab:data}). 
%More importantly, the difference of these factors between the cases and controls is negligible (see Fig.~\ref{fig:data})\cm{don't understand this sentence}.\jh{I think this sentence can be cut.}
This is different from some prior studies in the literature, in which the selection of the data is unclear and lacking a cohort diagram~\cite{laguarta2020covid, pinkas2020sars}.  
More importantly, we further performed experimental analysis to explore the effect of demographic bias on the model.
%\cm{this sentence seems to refer only to our first analysis and not to the exploration you do on gender bias?}\jh{added}

{\em Language Bias:} With the potential of COVID-19 digital testing to be applicable worldwide, it is important to explore the effect of language bias on different audio-based data (such as cough, breathing and voice).
%With applications for crowdsourcing this type of data gathering it worldwide, languages might add to the bias issue when trying to build generalizable machine learning models for audio~\cite{sagha2016cross}.
% \cm{can we cide something?}
%While our model takes voice singles\cm{singles? what do you mean?} as input, language might be another possible factor as our data were collected worldwide and with different languages. Also, 
%In a preliminary study, where language was not controlled, we found that the  trained model performed differently across participants using different languages, especially when the prevalence rate was  different across various geographical areas~\cite{Brown20-Exploring, Han2021exploring}. \cm{is it worth saying this given now we have our testing? I would cut this and just explain we have our model mainly tested on just engish and then test on the italians and say what we found eg effect on voice and not so much cough} \eb{agree}
To disentangle possible confounding effect of language, we restricted our analysis to English-speaking data, which gives the most realistic perspective of the capabilities of audio based diagnostics for COVID-19. 
In addition, similar to demographic bias, we carried out experiments to test the effect of language bias when the model was trained with unbalanced multi-language data.
%\cm{can we add something about the language bias test you have done?}\jh{added, please check}
%\cm{perhaps we can say that we have not explored yet the effect of accent and this could also have an effect}\jh{the effect of accent is not discussed in the limitation of our study. I think it fits there more?}

{\em User Independence:} Moreover, in some  prior studies, cross-validation was applied for performance validation: this is generally done when the data is scarce and user samples become very important. Data from the same participant might  be used for both model training and validation~\cite{xue2021exploring,pahar2021machine}:  while this might be considered acceptable in testing theoretical machine learning techniques,
if a user appears in both training and testing sets, such models typically do not generalise well, making them poorly-suited for a realistic setting.
%in realistic setting it would not be so common that the model could learn features from a user appearing both in training and testing. \eb{the second part of the sentence doesn't make much sense to me, should we just say that if a user appears in both training and testing sets, such models typically do not generalise well, making them poorly-suited for a realistic setting.}
With the luxury of a large dataset, we could choose to perform user-independent validation, where participants' data used for model validation are not included and unseen during model training. We are confident that this is a more realistic approach, which could inform future in-the-wild audio-based screening.

\vspace{.2cm}\noindent\textbf{Limitations of our study.}
Several limitations to our work should be acknowledged.
%   \tx{from the published nature digital medicine COVID-19 paper, they just have a short paragraph for limitation, and most are about data, not the design the study or model.}
COVID-19 is known to often manifest as respiratory symptoms, which are also common for other relatively widespread diseases, as well as among the smoking population. Therefore, we conducted an in-depth analysis to establish whether our model could be influenced by other respiratory pathology.  Specifically, we evaluated the ability of our model to correctly identify a COVID-19 infection in participants who indicated Asthma and High blood pressure in their medical history as well (as these are reasonably large cohorts in our data collection) compared to a cohort who indicated no other medical conditions. We also tested the model on participants with a variety of smoking status reported (e.\,g., few to many cigarettes per day). 
% Reasonable results indicate that a COVID-19 sound is likely to significantly differ from sounds generated by other respiratory or otherwise confounding conditions, and our model successfully captures features specific to COVID-19.
However, we note that we have not had the opportunity to test against a wider variety of specific respiratory infections, such as influenza or rhinovirus, since they were not prevalent when our data was collected and are difficult to have a reliable ground truth for.
%since we did not collect this information from the participants. 
It is possible, however, that the participants who reported a cough and had a negative COVID-19 test result were indeed suffering from some respiratory condition at the time of the sample collection. 

Also, as our models did not fully control for all potential confounding factors such as race and have much less number of elderly participants, future studies should investigate these biases. %\eb{this feels a little out of the blue, we talk about respiratory conditions and smoking, and then all of a sudden in the same paragraph start talking about race and age? I would move this to a separate, even if short, paragraph} 
In addition, though in the present study the language was well controlled (all English), it is yet unclear whether and how different types of accents would affect the model, while we lack such information to study this.
%\cm{talk about language accent?}\jh{added, pls check if fit} 
% With respect to the comparison with expert performance, it is worth noting that the model's predictions sometimes differed from those of the doctors. An example is the two asthmatic participants without any symptoms (in Extended Data Table~\ref{tab:8users}, sample 5 and 6). While three to five of the seven doctors voted for COVID-19 positive based on the given audio materials, showing difficulty in distinguishing asthma from COVID-19 on audio alone by respiratory doctors, the model generated correct estimations. In contrast, the model made incorrect predictions for another two participants (in Extended Data Table~\ref{tab:8users}, sample 7 and 8), where all doctors agreed and predicted these samples as COVID-19 positive. It seems that the model and human doctors may pick up on different aspects of the sounds for their decisions, and so our model could potentially serve as decision support for medical professionals.
 
Our data is  crowdsourced: we rely on the trustworthiness of the responses from individuals, especially with respect to their COVID-19 testing status. The scale of the data helps in amortising the noise generated by the crowdsourcing process while, at the same time, shows robustness of the approach to uncontrolled conditions. Our data, while aims to match the cohort to target population as much as possible, lacks clinical validation. Thus, additional external validation should be performed to assess the generalisation of the prediction model before being  applied in clinical practice.

% \jh{the following needs to revise}
% The performance show that sensitivity remains quite high (74\% generally), i.\,e., we are able to identify a high proportion of COVID-19 positive participants, however, achieving more than this would compromise specificity and result in many false positives: this is likely due to the noise inherent in the crowdsourced nature of the data.

% \tx{added:} 
\textbf{Potential implications for practice.} The model's trainable parameters are optimised based on a default threshold of 0.5 from the final softmax output layer (see Fig.~\ref{fig:framework}): this value is used to classify COVID-19 positive and negative predictions. While our model could be used on the general population for COVID-19 digital testing, we explore different contexts of applications where this threshold, as a hyper-parameter, can be adjusted for a more optional optimal outcome: we report the ROC curve and sensitivity/specificity under different decision thresholds for asymptomatic and symptomatic groups (participants who did and did not declare symptoms)
% (results for default threshold of 0.5 can be found Fig.~\ref{fig:results}) 
in \hyperlink{Appendix}{Appendix}: Fig.~\ref{fig:results_as} and Fig.~\ref{fig:results_ss}, respectively.   
Specifically, when applying the model with the aim of screening the asymptomatic population for risk of exposure, from Fig.~\ref{fig:results_as}b), a lower threshold can be used to guarantee a higher Youden Index (defined as Sensitivity + Specificity $- 1$) and a higher sensitivity compared to the threshold of 0.5, so that potential COVID-19 infections are exhaustively covered, and false positives can be easily filtered by a further clinical testing. Yet, if the targeted group is symptomatic (Fig.~\ref{fig:results_ss}b), to limit the false positives,
a higher specificity can be achieved by by slightly increasing the threshold to maximise the Youden Index. In this study, as we have limited samples for validation in our dataset, we only demonstrate the performance on the test set data under different threshold settings as a proof-of-concept. For clinical use, a further investigation is required on how to adjust and calibrate this threshold to meet different testing criteria. 

Finally, audio-based predictive models could be combined with other signatures from other biological signals such as heart rate~\cite{quer2021wearable}, as well as self-reported symptoms~\cite{Menni20-Real} for improved accuracy, however this would require crowdsourcing additional data from the participants. 
% \tx{why mentioned heart-rate but not symptom?}\jh{good point, revised by adding symptoms}

%as a standalone COVID-19 testing tool, further studies to study how to integrate it with other signatures from other biological signals such as heart rate~\cite{quer2021wearable} would be of great potential to facilitate remote monitoring and testing.

%\cm{can we cut this part?}
\textbf{Conclusion.} In this work, we have developed and validated a deep learning method for detecting COVID-19 solely by analysing human sounds via mobile or web applications. 
In particular, the crowdsourced data have been collected and processed to make the results  reliable, by controlling potential confounding factors in COVID-19 positive and negative cases. 
% We have evaluated this approach and studied the model performance using different metrics, including AUC-ROC, sensitivity, and specificity, on a dataset of 2,478 participants worldwide.
% , and conducted longitudinal analysis on another 71 positive participants who consistently contributed more than five samples. 
We analysed the presented model's predictive performance on detecting COVID-19 infection,
% \eb{I think this might not be the most appropriate use of "predicitive performance". Could say "We analysed (orcompared) model's predictive performance on detecting COVID-19 infection"}, 
which may bring insights into the adoption of digital health technologies in the COVID-19 era. 
Moreover, we analysed the risks of modelling with various biased data, which led to overestimated performance. This demonstrated that biased data or modelling should be avoided to rigorously validate the digital testing tool for clinical efficacy.
%\jc{no mention of bias etc. which is an important part of this study}\cm{yes this does not seem to appear to have been adapted after the recent results. perhaps cut as it does not add much}\jh{revised, pls check}
% The presented approach shows potential in a variety of use cases, including as an automated screening tool, providing complementary information for clinicians, and as a personalised tool to remotely monitor progression of disease.

\section*{\hypertarget{methodSection}{Methods}}\label{methods}

\begin{figure*}[t]
\centering
\includegraphics[width=0.9\textwidth]{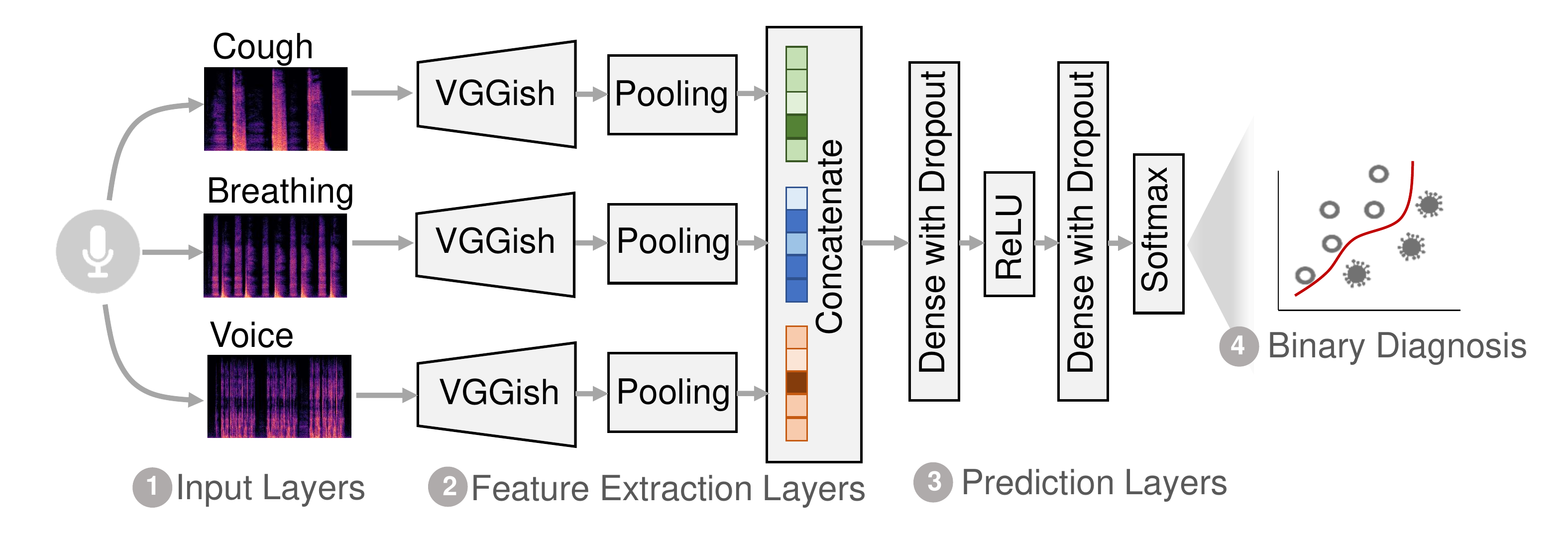}
\caption{\textbf{Overview architecture of the deep learning model}. \textnormal{A convolutional neural network using cough, breathing, and voice sounds as input, to predict COVID-19 as a binary outcome. This model can also be applied to longitudinal data iteratively to assess the clinical progression over time. (\textit{VGGish} is a neural network pre-trained on the Audioset dataset, \textit{Pooling} is an aggregation operator,  \textit{Dense} is a fully connected neural network layer, \textit{Dropout} is a randomized operation that reduces overfitting, \textit{ReLU} is a rectified linear unit activation, \textit{Softmax} is the logistic function.)} 
% \tx{Not sure whether this should go to appendix}
}
% \jh{remove part 5 and revise capitation accordlingly} \tx{done}}
\label{fig:framework}
\end{figure*}

% \jh{will it be good to move the experiment design to front as study goal?} \tx{Yes, I agree.}\jh{done}
% \tx{rephrase this section a lot. but need highlight some design in Intro, including well-controlled experiments and transfers learning technique?}

% \vspace{.2cm}\noindent\textbf{Study goals.}
% The primary goal of this study is to build a predictive model for COVID-19 detection via sound, in hopes that an automated audio-based COVID-19 digital testing tool will help achieve rapid, cost-effective, non-invasive COVID-19 testing. For this aim, we develop a predictive model which takes three sounds as predictors (breathing, cough sounds, and voice) and analyse its performance on a retrospective cohort. \tx{why using three modality here is not well connected. Maybe this part is not necessary?}

\vspace{.2cm}\noindent\textbf{Data collection and preparation.}
Our data were crowdsourced via a data gathering framework released in April 2020, in multiple languages and for multiple platforms (a webpage, an Android app, and an iOS app). Collected data consist of participants' age, gender, medical history, current symptoms, and three audio recordings: three voluntary cough sounds, three to five inhalation-exhalation sounds, and the participant reading a standard sentence from the screen three times. Participants were asked whether they had been tested for COVID-19, and an optional geo-location sample was collected. The mobile apps also prompted the participant to input symptoms and sounds every two days. %, allowing the study of changes in an individual’s respiratory sounds over time. 
No identifiable information was collected. 
% Collection and study of these data have been approved by the Ethics Committee of the Department of the Computer Science and Technology at the University of Cambridge.
As of 26th April 2021, a total of 36,364 participants contributed 75,201 samples to our project.
 
We used samples with self-reported COVID-19 test results for experiments as ground truth. Hence, 61,615 samples without reported test results were excluded. Further 110 samples with COVID-19 testing results declared to be obtained 2 weeks before the recordings were made were also discarded due to the delayed audio recordings with respect to COVID-19 testing.
%to guarantee the effectiveness of taking testing results as the ground-truth of our study. 
Our data was sourced in multiple languages (English, Italian, Spanish, Portuguese, etc.) and the number of samples in each language varied. To avoid language bias, for the main results of this paper we used English audio samples only, with 8,102 non-English samples excluded. Lastly, we manually checked the quality of each recording, deleting in total another 134 samples, that were either incomplete with recordings shorter than 2 seconds, or samples with silent recordings, or distorted samples with poor audio quality. As a result, 5,240 samples from 2,478 participants were explored for the majority of the experiments.
%\cm{do we want to say we use other data for the biases exploration?}

\vspace{.2cm}\noindent\textbf{Data used and experiment design for bias evaluation.}
%\cm{should this be moved up to my previous note?}
In addition to the above-mentioned data, we also prepared four datasets with known biases to validate the impact of confounding factors on audio based COVID-19 testing, by selecting from all eligible samples with COVID-19 test results. 
To be more specific, the following strategies were followed to generate the data:
\begin{itemize}
\item \textbf{Splitting:} Our balanced training set and testing set contained 1000 participants (800 participants for training\&validation, 200 for testing) and 1486 samples (1162 samples for training\&validation, and 329 for testing), with 1.5 samples per participant on average.  Instead of splitting training and testing by participants, for this comparison group, we randomly shuffled all samples and split them into training and testing according to the original ratio (1162:329).   
\item \textbf{Gender bias:} 
%Sometimes data collection produces may include bias due to many reasons, and one common bias lies in the gender.
To simulate the scenario where COVID-19 positive rate is significantly different in different gender groups,  which raises the concern that the model is detecting gender instead of COVID-19,  we manually selected 500 positive and 500 negative participants from the total 2,478 participants (blue box in Fig.~\ref{fig:data}) with gender distribution bias. Specifically, 56\% of the positive group are male and the rest 44\% are female, while in the negative group, females account for 85\% and males for \%15. Age demographics are kept balanced  and the total number of participants is unchanged, as shown in \hyperlink{Appendix}{Appendix} Fig.~\ref{fig:results_age}.
%\cm{unclear you have 2K participants, take 85 percent of the female group (perhpas mention how many these is out of 2k) and do what with the positives?} \jc{On this note please check each number/stats we report in table, text, figures. we cannot be wrong here.}%As such, gender bias was introduced and the impact can be tested. 
\item \textbf{Age bias:} With the same approach, for negative participants, we also purposefully selected 1) those aged over 39, and 2) those aged under 39 to simulate the scenarios when participants were not from the whole population. The revised distribution can be found in \hyperlink{Appendix}{Appendix} Fig.~\ref{fig:results_gender}.
\item \textbf{Language bias:} Rather than using all English speakers, to investigate the effect of language, 
%varied COVID-19 prevalence rates across the world, 
we replaced some English-speaking participants with Italian-speaking participants.  Specifically, we used more positively-tested Italians than negative. As a result, the positive group mainly consists of Italian-speakers, indicating the bias that participants who speak Italian are more likely to be COVID-19 positive. The detailed percentage can be found in \hyperlink{Appendix}{Appendix} Fig.~\ref{fig:results_lan}.
\end{itemize}

\vspace{.2cm}\noindent\textbf{Data processing.} 
For data pre-processing, all the collected audio recordings were resampled to \SI{16}{\kilo\hertz} and converted to mono channel. Then, these audio recordings were cropped by removing the silence periods at the beginning and the end of the recording, after which each sample was normalised.
%\eb{I think this paragraph is out of place, perahps should be moved to "data collection and preparation"? or have its own mini section}

\vspace{.2cm}\noindent\textbf{Model architecture.} 
We implemented a Convolutional Neural Network (CNN) based model for COVID-19 classification, as shown in \hyperlink{Appendix}{Appendix} Fig.~\ref{fig:framework}.
The network receives one sample with three audio recordings as input, including breathing, cough, and voice from one participant. A spectrogram is computed for each of the recordings and is fed into a \textit{VGGish} subnetwork. \textit{VGGish} is a state-of-the-art pre-trained CNN, with which we leverage and transfer the knowledge learnt from an external massive general-audio dataset~\cite{hershey2017cnn}. Each \textit{VGGish} block transforms the input spectrogram into a latent feature vector, then the features from three sound types are concatenated, and finally fed into a binary classifier. Such model design allowed the three sound types to be analysed jointly. 
Specifically, the model is composed of three parts (Fig.~\ref{fig:framework}):

{(1) Input Layers:} 
%For each audio recording, a Mel spectrogram is computed using magnitudes of the Short-Time Fourier Transform with a window size of \SI{25}{\milli\second}, a window hop of \SI{10}{\milli\second}, and a periodic Hann window. It is mapped by 64 mel bins covering the frequency range from \SIrange{125}{7500}{\hertz}. A log is taken from the resulting spectrogram with an offset, and the features are then framed into non-overlapping instances of \SI{0.96}{seconds}, where each one covers 64 mel bands and 96 frames of \SI{0.25}{million seconds} \SI{10}{\milli\second}.
The audio recording is first chunked into non-overlapping segments of 0.96 seconds. Log-mel spectrogram is computed for each segment with a window size of \SI{25}{ms}, a window hop of \SI{10}{ms}, and a periodic Hanning window. 64 Mel bins are adopted for Mel spectrogram covering the frequency range from \SI{125}{\hertz} to \SI{7500}{\hertz}.%\td{Hz not showing?}. 
A small offset is used to convert the mel-spectrogram into log-scale, resulting in a log-mel spectrogram with size of 64 x 96 per chunk. 
%\td{Hamming or Hanning?} \tx{For this paragraph, all yes.}

{(2) Feature Extraction Layers:} The main component of the model is VGGish, a CNN-based network with cascaded convolutional layers, max-pooling,  and fully connected layers. This network transforms each input spectrogram frame into a 128-dimensional feature vector. Then, an average pooling layer is employed to aggregate all frames within one audio recording into one fixed-length latent feature vector. The size of the CNN kernels and the number of hidden states of fully connected layers are kept consistent with the original work~\cite{hershey2017cnn}.
%\td{The CNN kernel size and the number of neurons for all layers?}\tx{not necessary because it is a standard model?}. 

{(3) Prediction Layers:} The resulting latent feature vectors for three modalities are concatenated, and fed into the binary classifier, which consists of two dense layers (the number of hidden states are 96 and 2, respectively) with non-linear ReLU and Softmax activation functions, respectively. The output of the model is a continuous score within the range of 0 to 1 (i.e., the probability of the participant being infected with COVID-19), which can then be categorised into a binary score (0: negative, and 1: positive) with a threshold of 0.5.

To improve the robustness and generalisation ability of our deep learning model, the following techniques were employed:
\begin{itemize}
\item Transfer learning: Our collected data is relatively small compared to the number of parameters in the proposed deep neural network. In light of this, we harness transfer learning to improve the representing ability. Specifically, VGGish layers are initialised by a pre-train model, which is designed for audio classification task. %\tx{Digital Medicine seems a venue for machine learning, shall we show the results without transfer learning, and feature+SVM, as baselines in appendix?}\jh{not necessary I think}
\item Differential learning rate: Both VGGish and dense layers are jointly updated by using  our audio data. However,  we used a small learning rate for parameter update of the VGGish part of the network, and increased the learning rate 10 times for the dense layers. Specifically, learning rate was set as 1e-6 for VGGish and 1e-5 for the final dense layers. 
\item Avoiding over-fitting: We utilised learning rate decay (factor = 0.9) and L2-regularisation (penalty coefficient = 1e-6).
\item Two-phase training: To use the data more efficiently, we primarily trained the model via training set and identified the best hyper-parameters based on the averaged sensitivity and specificity of the 15$^\text{th}$ epoch on validation set,  and then we merged the training set to fine tune the model until the training performance kept unchanged.
\end{itemize}

Parameters of the deep learning model were updated by iterative gradient back propagation by the binary cross-entropy loss function on training set. The training batch size was 1. The whole framework was implemented by Python 3.6 and Tensorflow 1.15. Model training was done on a Nvidia Quadro RTX 8000 GPU.% and result analysis was done on a 64GB CPU.
%\jc{1 is  it correct? Why? A batch size less than 8 is known to perform not well. Standard is 32}

\vspace{.2cm}\noindent\textbf{Experiment design.}
% Such model has a large number of parameters to estimate, and hence requires a learning process to find the optimal values.
We performed participant-independent splitting, which means that samples from a participant included in training set for parameters estimation will not be used for model evaluation.%\cm{do we want to say this is more realistic?} \tx{yes}
We randomly selected 80\% of all positive participants for model learning and sampled the same number of negative participants by maintaining a similar demographic distributions in these two groups (see \hyperlink{Appendix}{Appendix} Table~\ref{tab:data}). This can minimise the data bias collected in a crowd-sourcing manner.
10\% of participants were held out for hyper-parameters searching like the size of dense layers. 
Once a final trained model was obtained, the performance were evaluated on different demographics, prevalence levels, and health conditions data. 
Measures of performance include the Area Under the Receiver Operating Characteristic Curve (AUC-ROC), sensitivity, and specificity. 
For all the metrics, we calculated 95\% confidence intervals (95\% CIs), using bootstrap re-sampling with 1000 bootstrap samples and replacement~\cite{Carpenter00-Bootstrap}.

\vspace{.2cm}\noindent\textbf{Data Availability}\\
The data is sensitive as voice sounds can be deanonymised. Anonymised data will be made available for academic research upon requests directed to the corresponding author. Institutions will need to sign a data transfer agreement with the University of Cambridge to obtain the data. 
A copy of the data will be transferred to the institution requesting the data. We already have the data transfer agreement in place.

\vspace{.2cm}\noindent\textbf{Ethics}\\ 
The study was approved by the ethics committee of the Department of Computer Science at the University of Cambridge, with ID \#722.
Our app displays a consent screen, where we ask the user’s permission to participate in the study by using the app. 
Also note that the legal basis for processing any personal data collected for this work is to perform a task in the public interest, namely academic research.
More information is available at \url{https://covid-19-sounds.org/en/privacy.html}.
% This study was approved by the NYU Langone Health Institutional Review Board (IRB), with ID# i20-00858. 
%Collection and study of these data have been approved by the Ethics Committee of the Department of the Computer Science and Technology at the University of Cambridge.
%A waiver for informed consent was granted by the ???, since the study presents no more than minimal risk.

\vspace{.2cm}\noindent\textbf{Code Availability}\\
Python code and parameters used for training of neural networks will be available on GitHub for reproducibility purposes.

\vspace{.2cm}\noindent\textbf{Acknowledgements}\\ 
This work was supported by ERC Project 833296 (EAR) and the
UK Cystic Fibrosis Trust. We thank everyone who volunteered
their data. 
%We are also grateful to Dr. L. Marques-Fernandez, F. Pamatat, T. Parcollet, J. Mante, G. Nettesheim, M. Tsitoura, and C. Manolache for helping with the app translations\jh{written consent of any cited individual noted in acknowledgments to be included}.

\vspace{.2cm}\noindent\textbf{Author contributions}\\ 
% 14 roles that can be attributed to authors: conceptualisation, data curation, formal analysis, funding acquisition, investigation, methodology, project administration, resources, software, supervision, validation, visualisation, writing – original draft, and writing– review & editing.
% Chloë Brown (CB), Jagmohan Chauhan (JC), Andreas Grammenos (AG), Jing Han (JH), Apinan Hasthanasombat (AH), Dimitris Spathis(DS), Tong Xia(TX), Pietro Cicuta(PC), Cecilia Mascolo(CM), Andres Floto(AF), Erika Bondareva (EB), Ting Dang (TD)
AF, CM, PC designed the study. 
% \jh{Todo for Ting:TD to be added somewhere :)} \td{Added}
AH, AG, CB, DS, JC designed and implemented the app to collect the sample data.
AG designed and implemented the server infrastructure.
%DS, EB, JH, TX, selected the data for analysis.
EB, JH, TX, selected the data for analysis.
TX developed the neural network models, conducted the experiments and generated all tables and figures in the manuscript. 
%AH, DS, JH, TX performed the statistical analysis. 
JH, TX performed the statistical analysis. 
CB, DS, EB, JH, TX, TD wrote the first draft of the manuscript.
All authors vouch for the data, analyses, and interpretations.
All authors critically reviewed, contributed to the preparation of the manuscript, and approved the final version.

% \noindent\textbf{Declaration of interests}\\ 
\vspace{.2cm}\noindent\textbf{Competing interests}\\ 
All authors declare no competing interests.

\bibliographystyle{vancouver}
\bibliography{cas-refs}

% \bibliographystyle{main}{plain}
% \bibliography{main}{cas-refs}{References}

% \clearpage
% \newbibstartnumber{29}
% \begin{bibunit}[plain]
%     \section*{Methods} 
%     \label{methods}
%     \input{methods}
%     % \bibliographystyle*{vancouver}
%     \putbib
% \end{bibunit}
% % \bibliographystyle{appendix}{vancouver}
% % \bibliography{appendix}{cas-refs}{References}

% \newpage

% \section{Bibliography styles}

% There are various bibliography styles available. You can select the
% style of your choice in the preamble of this document. These styles are
% Elsevier styles based on standard styles like Harvard and Vancouver.
% Please use Bib\TeX\ to generate your bibliography and include DOIs
% whenever available.

% Here are two sample references: 
% \cite{Fortunato2010}
% \cite{Fortunato2010,NewmanGirvan2004}
% \cite{Fortunato2010,Vehlowetal2013}

% \appendix
% \section{My Appendix}
% Appendix sections are coded under \verb+\appendix+.

% \verb+\printcredits+ command is used after appendix sections to list 
% author credit taxonomy contribution roles tagged using \verb+\credit+ 
% in frontmatter.


%\clearpage
% \balance
% \bibliographystyle{cas-model2-names}
% \bibliographystyle{ieeetr}
% \bibliographystyle{unsrtnat}

% %\vskip3pt

% \clearpage
\newpage
\clearpage
\appendix
\onecolumn
\setcounter{figure}{0}

\clearpage
\section*{\hypertarget{Appendix}{Appendix}}\label{Appendix}
\setcounter{table}{0}
\label{Supporting}
\addcontentsline{toc}{section}{Appendices}
\renewcommand{\thesubsection}{\Alph{subsection}}

\begin{table*}[h]
\centering
\renewcommand*{\tablename}{Table}
\caption{Demographics distribution of the training and balanced testing set.}
\resizebox{0.95\textwidth}{!}{
\begin{tabular}{p{2cm}cccccc}
\toprule
       & \multicolumn{3}{c}{Positive}     & \multicolumn{3}{c}{Negative} \\ \cmidrule(r){2-4} \cmidrule(r){5-7}  
& Training      & Validation  & Testing  & Training   & Validation   & Testing  \\ \hline
Male            &192(54.9\%)&29(58.0\%)&58(58.0\%) )&188(53.7\%)&31(62.0\%)&52(52.0\%)  \\ 
Female          &154(44.0\%)&20(40.0\%)&42(42.0\%)  &159(45.4\%)&19(38.0\%)&46(46.0\%) \\ 
16-29           &92(26.3\%)&10(20.0\%)&22(22.0\%)   &73(20.9\%)&16(32.0\%)&21(21.0\%)\\
30-39           &94(26.9\%)&16(32.0\%)&33(33.0\%)   &97(27.7\%)&13(26.0\%)&33(33.0\%)\\
40-49           &86(24.6\%)&13(26.0\%)&23(23.0\%)   &86(24.6\%)&10(20.0\%)&24(24.0\%)\\
50-59           &39(11.1\%)&5(10.0\%)&13(13.0\%)    &50(14.3\%)&4(8.0\%)&10(10.0\%)\\
60-69           &17(4.9\%) &1(2.0\%)&3(3.0\%)       &18(5.1\%)&4(8.0\%)&4(4.0\%)\\
70-             &6(1.7\%)  &2(4.0\%)&1(1.0\%)       &7(2.0\%)&1(2.0\%)&2(2.0\%)   \\
\bottomrule
\end{tabular}}
\justify A small proportion (under 5\%) of the participants in each set preferred not to answer about their age or gender.
\label{tab:data}
\end{table*}

\newpage
\begin{figure*}[t]%[pos=t,width=3cm,align=\centering]
\centering
{\includegraphics[width=0.99\textwidth]{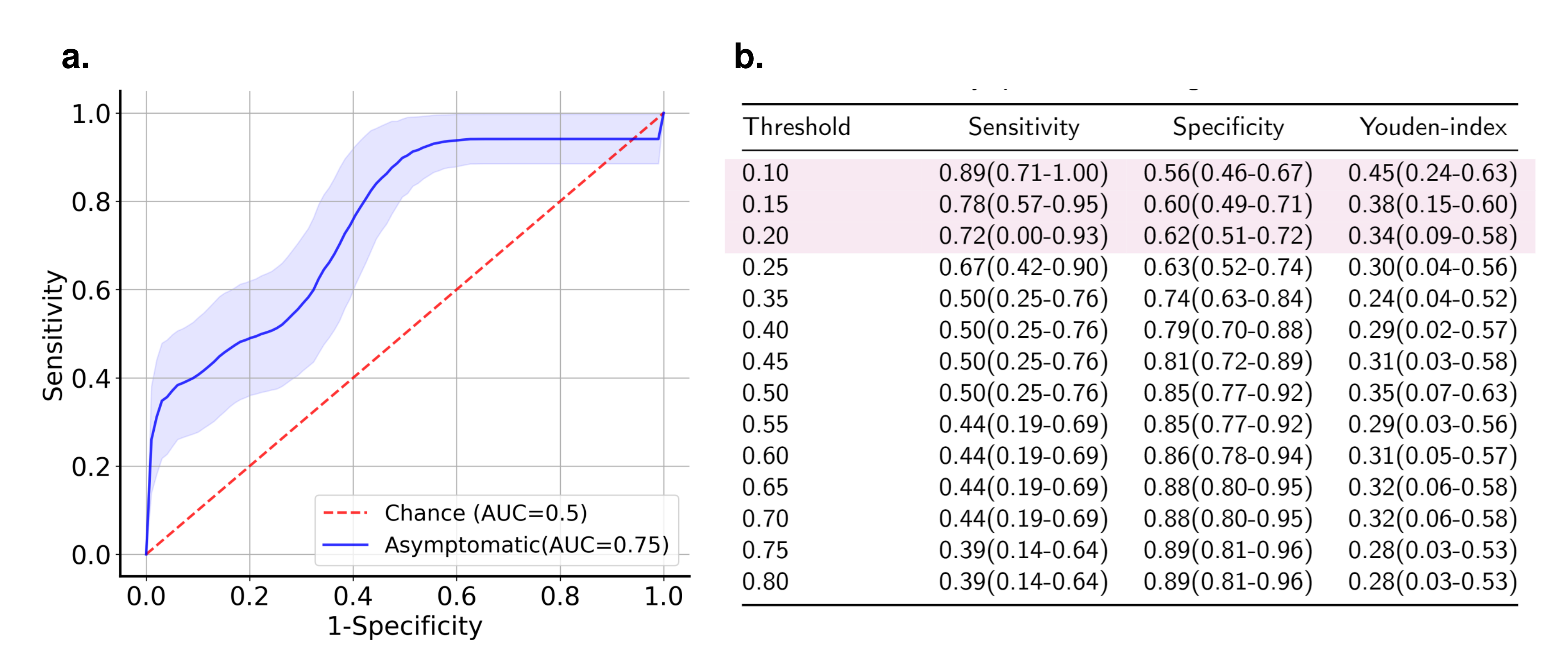}}
\caption{\textbf{Model performance for asymptomatic screening}. \textnormal{\textbf{a,} Receiver-operating characteristic curve for the binary classification task of diagnosing COVID-19. \textbf{b,} Sensitivity and specificity with 95\% CIs under different thresholds. The default value is 0.50.}}
\label{fig:results_as}
\end{figure*}

\begin{figure*}[t]%[pos=t,width=3cm,align=\centering]
\centering
{\includegraphics[width=0.99\textwidth]{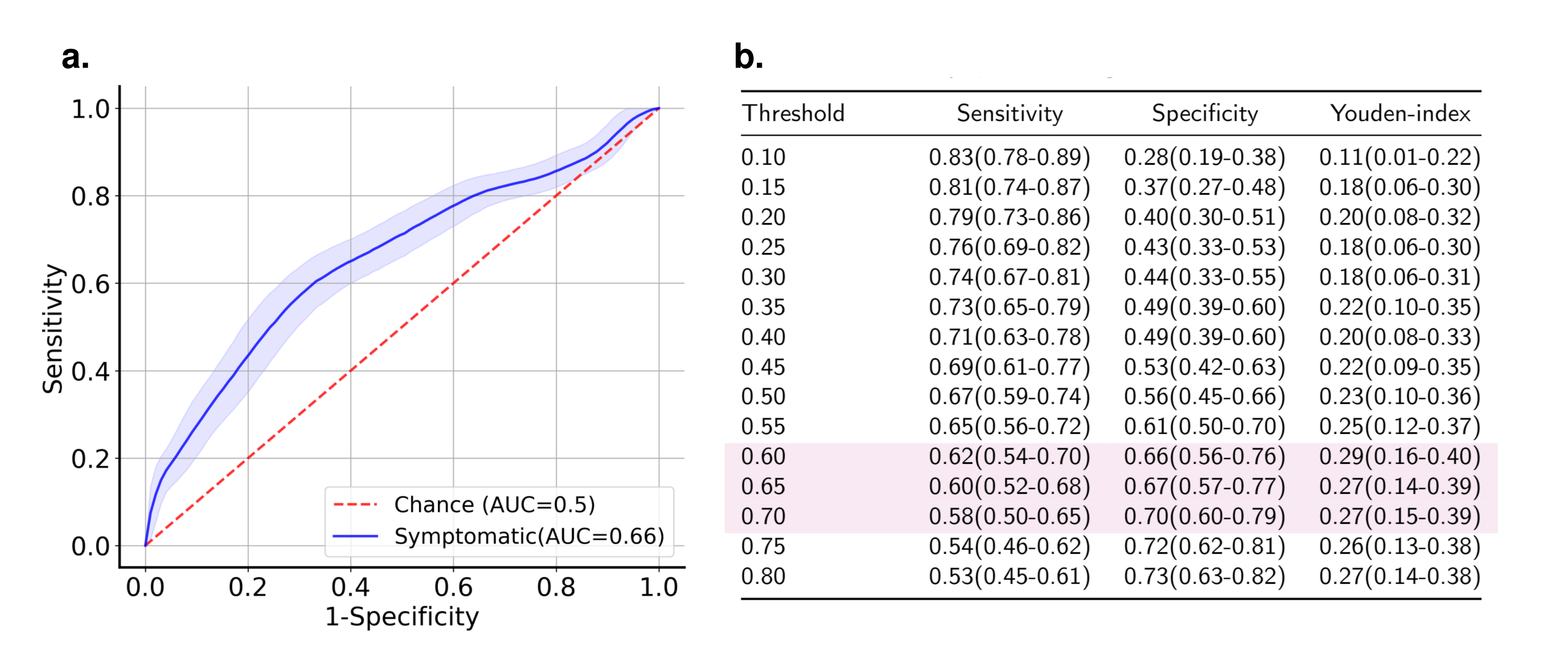}}
\caption{\textbf{Model performance for symptomatic diagnosis}. \textnormal{\textbf{a,} Receiver-operating characteristic curve for the binary classification task of diagnosing COVID-19. \textbf{b,} Sensitivity and specificity with 95\% CIs under different thresholds. The default value is 0.50.}}
\label{fig:results_ss}
\end{figure*}

\clearpage
\definecolor{LightCyan}{rgb}{0.8,1,1}

\begin{table}[h]
\centering
\caption{Performance of subgroups based on the model trained with \textbf{Random Splits}. }
\resizebox{0.75\textwidth}{!}{
\begin{tabular}{@{}p{1.5cm}ccccc@{}}
\toprule
Subgroup  & \#Users(\#Samples) &  AUC(95\% CI)  & Sensitivity(95\% CI) & Specificity(95\% CI)  \\ \hline
\textbf{Total}   &143(161)/139(158)&0.80(0.75-0.84)&0.71(0.63-0.77)&0.74(0.68-0.81)\\ \hline
\textbf{Gender}    &                                                        \\
Male              &76(85)/77(85)&0.81(0.74-0.87)&0.72(0.62-0.81)&0.75(0.66-0.84)  \\
Female            &65(74)/62(73)&0.79(0.71-0.86)&0.70(0.61-0.81)&0.73(0.62-0.83) \\  \hline
\textbf{Age}                                                \\
16-39             &69(79)/59(63)&0.82(0.75-0.89)&0.73(0.63-0.83)&0.76(0.65-0.87)   \\ 
40-59             &59(65)/57(70)&0.76(0.68-0.84)&0.66(0.54-0.77)&0.70(0.59-0.81) \\
60-               &10(12)/15(16)&0.86(0.71-0.98)&0.83(0.58-1.0)&0.88(0.69-1.0) \\    
% \hline
% \textbf{Symptom}    &                                                        \\
% Asmptomatic      &20(20)/73(81)&0.75(0.63-0.88)&0.7(0.5-0.89)&0.73(0.64-0.82)\\
% Symptomatic     &125(141)/70(77)&0.8(0.74-0.86)&0.71(0.63-0.79)&0.75(0.66-0.85) \\ \hline
\textbf{Overlap}    &                                                        \\
\rowcolor{LightCyan} Seen    &50(62)/59(76)&0.90(0.85-0.95)&0.84(0.75-0.92)&0.78(0.68-0.87) \\
                  Unseen     &93(99)/80(82)&0.70(0.62-0.77)&0.63(0.53-0.72)&0.71(0.60-0.81)\\  
\bottomrule
\end{tabular}}
\label{tab:UserBias}
\vspace{-0.1cm}
\end{table}

\begin{figure*}[h]%[pos=t,width=3cm,align=\centering]
\centering
{\includegraphics[width=0.75\textwidth]{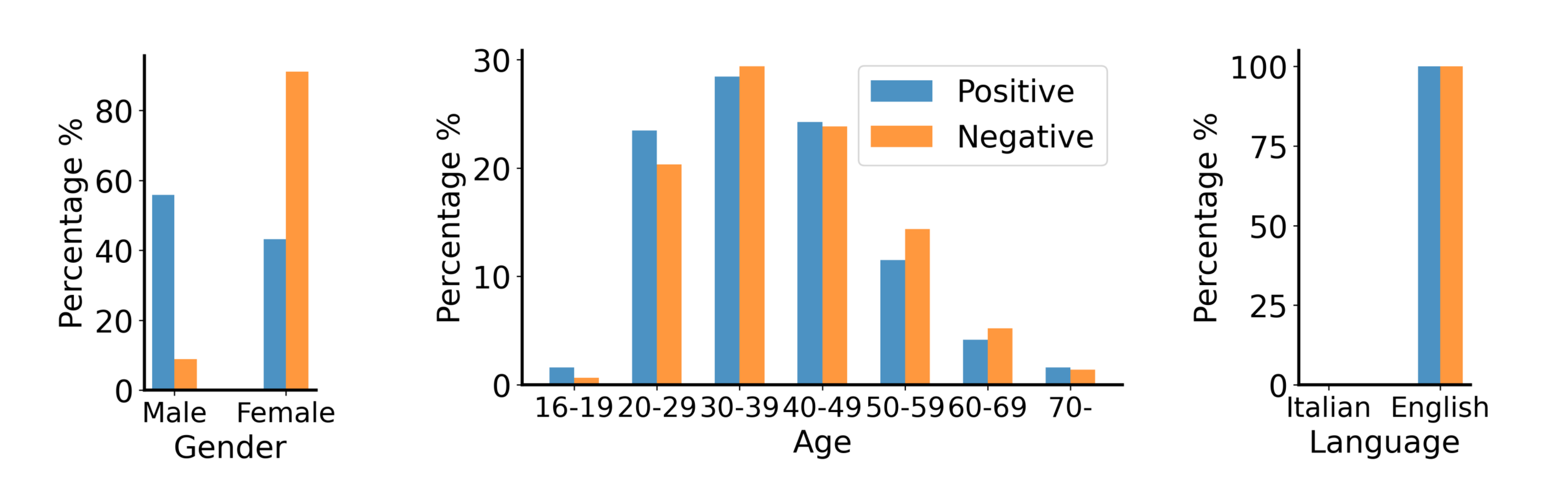}}
\vspace{-.3cm}
\caption{\textbf{Overall Demographic distribution fro gender bias evaluation.}}
\label{fig:results_gender}
\vspace{-.3cm}
\end{figure*}

\begin{table}[h]
\centering
\caption{Performance of subgroups based on the model trained with \textbf{Gender Bias}. }
\resizebox{0.75\textwidth}{!}{
\begin{tabular}{@{}p{1.5cm}ccccc@{}}
\toprule
Subgroup  & \#Users(\#Samples) &  AUC(95\% CI)  & Sensitivity(95\% CI) & Specificity(95\% CI)  \\ \hline
\textbf{Total}   &100(162)/100(187)&0.75(0.70-0.80)&0.46(0.38-0.54)&0.90(0.86-0.95)\\ \hline
\textbf{Gender}    &                                                        \\
 Male                       &58(85)/9(18)  &0.62(0.49-0.74)&0.66(0.56-0.76)&0.61(0.36-0.83) \\
\rowcolor{LightCyan} Female &42(77)/91(169)&0.66(0.58-0.74)&0.23(0.14-0.33)&0.93(0.90-0.97)\\  \hline
\textbf{Age}                                                \\
16-39              &55(81)/53(83)&0.69(0.61-0.76)&0.41(0.30-0.52)&0.84(0.76-0.92)   \\ 
40-59              &36(68)/38(86)&0.79(0.71-0.85)&0.56(0.44-0.67)&0.94(0.89-0.99) \\
60-                &4(8)/7(16)&0.96(0.86-1.00)&0.25(0.00-0.62)&1.00(1.00-1.00) \\    
% \hline
% \textbf{Symptom}    &                                                        \\
% Asymptomatic      &14(18)/44(93)&0.77(0.63-0.89)&0.39(0.15-0.62)&0.96(0.91-0.99)\\   
% symptomatic     &92(144)/68(94)&0.7(0.63-0.77)&0.47(0.38-0.55)&0.85(0.78-0.92) \\
\bottomrule
\end{tabular}}
\label{tab:GenderBias}
\end{table}

\clearpage

\begin{figure*}[h]%[pos=t,width=3cm,align=\centering]
\centering
{\includegraphics[width=0.95\textwidth]{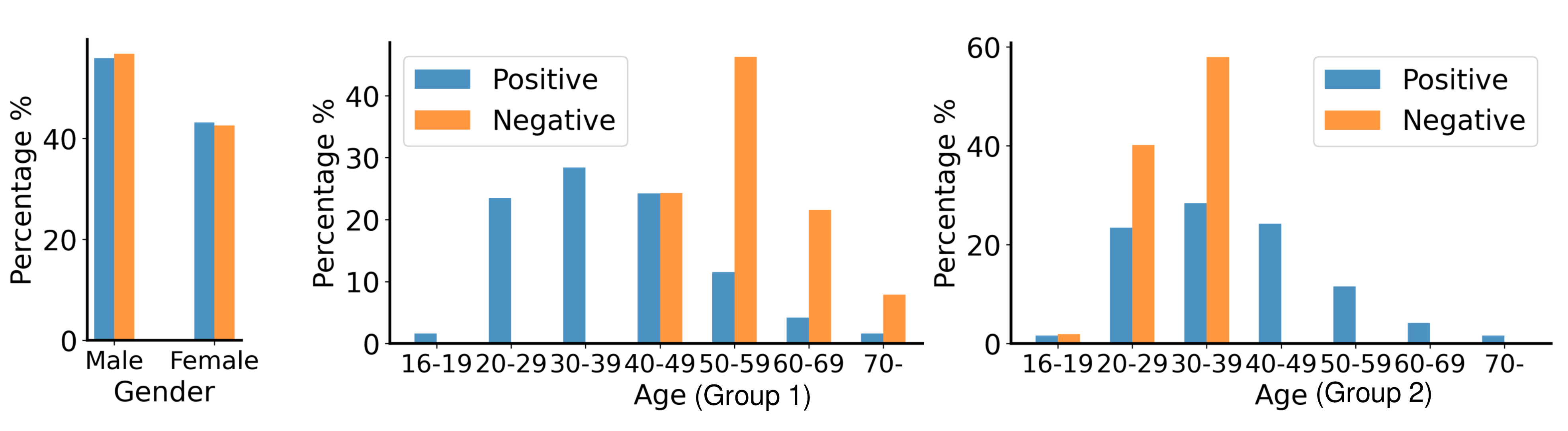}}
\vspace{-.3cm}
\caption{\textbf{Overall Demographic distribution fro age bias evaluation.}}
\label{fig:results_age}
\vspace{-.3cm}
\end{figure*}

\begin{table}[h]
\centering
\caption{Performance of subgroups based on the model trained with \textbf{Age Bias (Group 1)}. }
\resizebox{0.75\textwidth}{!}{
\begin{tabular}{@{}p{1.5cm}ccccc@{}}
\toprule
Subgroup  & \#Users(\#Samples) &  AUC(95\% CI)  & Sensitivity(95\% CI) & Specificity(95\% CI)  \\ \hline
\textbf{Total}  &100(162)/100(162)&0.80(0.74-0.84)&0.71(0.63-0.78)&0.73(0.66-0.79)\\ \hline
\textbf{Gender}    &                                                        \\
 Male   &58(85)/58(97)&0.81(0.75-0.87)&0.71(0.61-0.80)&0.77(0.68-0.85)\\
 Female &42(77)/42(65)&0.78(0.71-0.85)&0.71(0.61-0.81)&0.66(0.55-0.78)\\  \hline
\textbf{Age}                                                \\
16-39              &55(81)/0(0)&- &0.75(0.65-0.84)&-\\ 
40-49              &23(41)/25(31)&0.70(0.58-0.81)&0.73(0.59-0.86)&0.61(0.44-0.78)\\
50-59              &13(27)/47(74)&0.74(0.61-0.86)&0.67(0.47-0.84)&0.73(0.63-0.83)\\
\rowcolor{LightCyan} 60- &4(8)/28(57)&0.62(0.45-0.79)&0.25(0.00-0.60)&0.79(0.68-0.89) \\    
% \hline
% \textbf{Symptom}    &                                                        \\
% Asymptomatic     &14(18)/50(81)&0.84(0.7-0.96)&0.78(0.56-0.95)&0.75(0.66-0.84)\\
% Symptomatic     &92(144)/59(81)&0.77(0.71-0.83)&0.7(0.62-0.77)&0.7(0.6-0.79) \\   
\bottomrule
\end{tabular}}
\label{tab:AgeBias}
\end{table}

\begin{table}[h]
\centering
\caption{Performance of subgroups based on the model trained with \textbf{Age Bias (Group 2)}. }
\resizebox{0.75\textwidth}{!}{
\begin{tabular}{@{}p{1.5cm}ccccc@{}}
\toprule
Subgroup  & \#Users(\#Samples) &  AUC(95\% CI)  & Sensitivity(95\% CI) & Specificity(95\% CI)  \\ \hline
\textbf{Total}  &100(162)/100(121)&0.65(0.58-0.71)&0.53(0.45-0.61)&0.69(0.61-0.78)\\ \hline
\textbf{Gender}    &                                                        \\
 Male  &58(85)/57(65)&0.64(0.55-0.72)&0.48(0.38-0.59)&0.69(0.58-0.80)\\
 Female &42(77)/43(56)&0.66(0.57-0.75)&0.58(0.48-0.69)&0.70(0.57-0.82)\\  \hline
\textbf{Age}                                                \\
\rowcolor{LightCyan}16-39              &55(81)/100(121)&0.60(0.52-0.68)&0.44(0.34-0.55)&0.69(0.60-0.77)\\ 
40- &40(76)/0(0)&-&0.63(0.53-0.74)&- \\    
% \hline
% \textbf{Symptom}    &                                                        \\
% Asymptomatic     &14(18)/35(41)&0.55(0.37-0.71)&0.28(0.08-0.5)&0.78(0.64-0.9)\\
% Symptomatic      &92(144)/70(80)&0.64(0.57-0.72)&0.56(0.49-0.65)&0.65(0.54-0.75)\\   
\bottomrule
\end{tabular}}
\label{tab:AgeBias}
\end{table}

\clearpage

\begin{figure*}[t]%[pos=t,width=3cm,align=\centering]
\centering
{\includegraphics[width=0.85\textwidth]{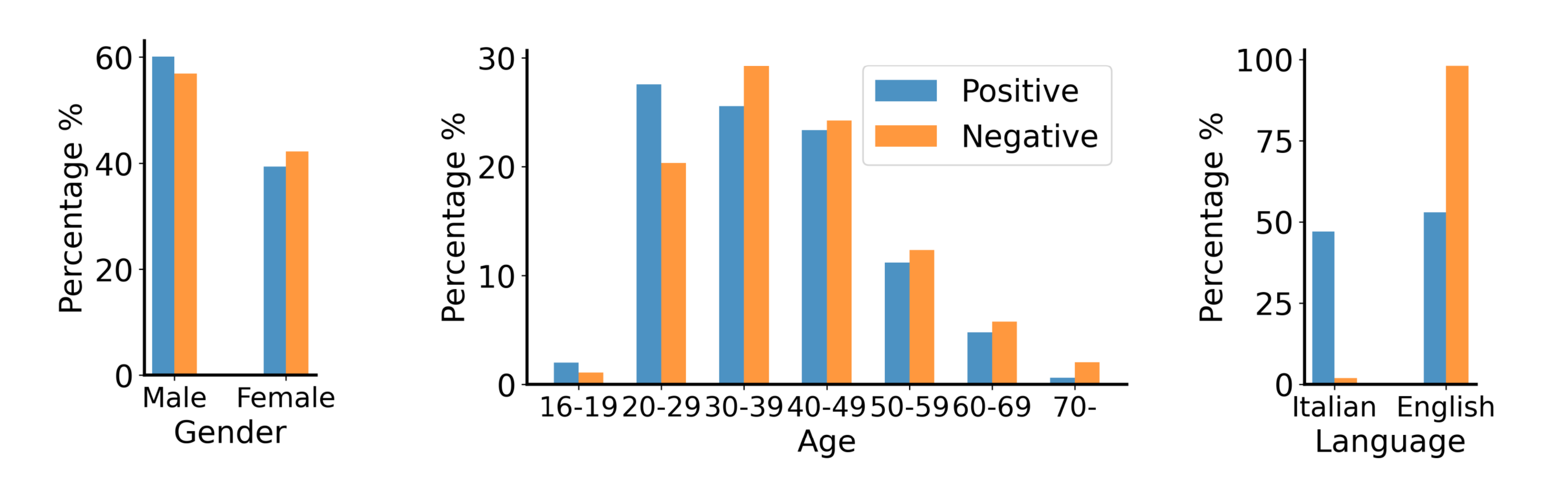}}
\caption{\textbf{Overall Demographic distribution for language bias evaluation.}}
\label{fig:results_lan}
\end{figure*}

\begin{table}[h]
\centering
\caption{Performance of subgroups based on the model trained with \textbf{Language Bias}. }
\resizebox{0.75\textwidth}{!}{
\begin{tabular}{@{}p{1.5cm}ccccc@{}}
\toprule
Modality  & \#Users(\#Samples) &  AUC(95\% CI)  & Sensitivity(95\% CI) & Specificity(95\% CI)  \\ \hline
\textbf{Three}  \\
Total   &100(139)/100(183)&0.77(0.71-0.82)&0.60(0.51-0.68)&0.78(0.72-0.84)\\ 
\rowcolor{LightCyan}English &52(73)/98(177)&0.61(0.54-0.69)&0.25(0.15-0.36)&0.81(0.75-0.86)\\
\rowcolor{LightCyan}Italian   &48(66)/2(6)&0.42(0.22-0.67)&0.98(0.95-1.0)&0.00(0.00-0.00)\\\hline
\textbf{Breathing}    &                                                        \\
Total &100(139)/100(183)&0.55(0.49-0.62)&0.56(0.48-0.65)&0.58(0.51-0.66)\\ 
English &52(73)/98(177)&0.53(0.45-0.61)&0.51(0.39-0.63)&0.60(0.53-0.67)\\
Italian   &48(66)/2(6)&0.26(0.06-0.55)&0.62(0.50-0.74)&0.17(0.00-0.57)\\\hline
\textbf{Cough}    &                                                        \\
 Total &100(139)/100(183)&0.64(0.58-0.7)&0.62(0.53-0.7)&0.59(0.51-0.66)\\ 
English &52(73)/98(177)&0.59(0.51-0.68)&0.55(0.44-0.67)&0.60(0.52-0.68)\\
Italian &48(66)/2(6)&0.57(0.27-0.81)&0.70(0.59-0.81)&0.33(0.00-0.75)\\\hline
\textbf{Voice}    &                                                        \\
Total &100(139)/100(183)&0.77(0.71-0.82)&0.59(0.50-0.67)&0.82(0.76-0.88)\\ 
\rowcolor{LightCyan}English &52(73)/98(177)&0.60(0.52-0.67)&0.22(0.12-0.32)&0.85(0.79-0.90)\\
Italian &48(66)/2(6)&0.64(0.49-0.77)&1.00(1.00-1.00)&0.00(0.00-0.00)\\
\bottomrule
\end{tabular}}
\label{tab:AgeBias}
\end{table}

\end{document}